\documentclass[11pt,twoside]{article}
\usepackage{times}
\usepackage[T1]{fontenc}
\usepackage{geometry}
\usepackage{graphics}

\makeatletter

\providecommand{\LyX}{L\kern-.1667em\lower.25em\hbox{Y}\kern-.125emX\@}

 \usepackage{verbatim}

\usepackage{latexsym}
\usepackage{amssymb}
\makeatother

\begin{document}

\noindent \markboth{THE BROAD BRILLOUIN DOUBLET AND CP OF KTaO$_3$}{E. FARHI \textit{et al.}} \thispagestyle{plain}

\title{ }

\noindent THE EXTRA BRILLOUIN DOUBLETS AND CENTRAL PEAK OF KTaO\( _{3} \) :
\\
SECOND SOUND \emph{vs.} TWO-PHONON DIFFERENCE SCATTERING

\author{ }



\bigskip{}
{\par\noindent \raggedright E. FARHI\( ^{a,b} \), A.K. TAGANTSEV\( ^{c} \),
B. HEHLEN\( ^{a} \), R. CURRAT\( ^{b} \), \\
L.A. BOATNER\( ^{d} \), and E. COURTENS\( ^{a} \)\par}

{\par\raggedright {\small \( ^{a} \) Laboratoire des Verres, Universit\'e de
Montpellier 2, F-34095 Montpellier, France}\small \par}

{\par\raggedright {\small \( ^{b} \) Institut Laue Langevin, BP 156, 38042 Grenoble
Cedex 9, France }\small \par}

{\par\raggedright {\small \( ^{c} \) Laboratoire de C\'eramique, EPFL, CH-1015
Lausanne, Switzerland}\small \par}

{\par\raggedright {\small \( ^{d} \) Solid State Division, Oak Ridge Nat. Lab.,
Oak Ridge, TN 37831-6056, USA}\small \par}

\begin{abstract}
\underbar{Abstract} \hspace{0.5cm}Low-\( T \) Brillouin spectra of the incipient
ferroelectric KTaO\( _{3} \) exhibit a broad central peak (CP), and additional
Brillouin doublets (BD), that can both be related to phonon-density fluctuations.
On the basis of new high-resolution neutron data obtained of low-lying phonon
branches, we analysed the phonon-kinetics mechanisms that are possibly the origin
of these unusual features. Firstly, transverse acoustic (TA) phonons whose normal
damping is \textit{faster} than the BD frequency can produce hydrodynamic second
sound. Secondly, two-phonon difference scattering from \emph{low damping} thermal
transverse phonons contribute to the spectra with either a sharp or a broader
doublet, depending on the phonon group velocity and anisotropy of dispersion
surfaces. The position of the observed sharp doublet is consistent with both
mechanisms, but a comparison of the computed and experimental anisotropies favours
the second process.
\bigskip{}

\underbar{Keywords}\hspace{0.5cm}quantum paraelectrics, inelastic neutron and
Brillouin scattering, second sound, two-phonon difference scattering, phonon-density-fluctuation
processes.
\end{abstract}

\section{\textsc{\underbar{\large INTRODUCTION}}\large }

Both KTaO\( _{3} \) and SrTiO\( _{3} \) perovskite-type crystals exhibit ferroelectric
transition-like features with decreasing temperature. Their dielectric constant,
\( \epsilon  \), increases as the energy \( \Omega _{sm} \) of the associated
Brillouin zone (BZ) centre transverse optic (TO) mode decreases following the
classical Curie-Weiss law. However, zero-point quantum fluctuations prevent
critical mode condensation and the crystal remains paraelectric down to the
lowest temperatures (\( T \)). Such a material is then called a quantum paraelectric
(QPE) \cite{Kurtz75,Muller79,Tosatti94}. The proximity of the soft TO and acoustic
modes near BZ centre might significantly modify low-\( T \) phonon kinetics.
The case of SrTiO\( _{3} \) is more complicated than KTaO\( _{3} \), since
SrTiO\( _{3} \) undergoes a well-known structural transition at \( T_{a}\sim 105 \)
K \cite{Unoki67,Thomas68,Shirane69}. Below this temperature, the unit cell
doubles and the crystal usually forms domains. In addition to this experimental
difficulty, the description becomes more involved due to the presence of three
additional low frequency branches. For this reason, we first investigate KTaO\( _{3} \),
which remains in a cubic \( Pm3m \) space group symmetry down to the lowest
temperatures. 

Many inelastic neutron-scattering \cite{Axe70,Comes72,Perry89}, Raman \cite{Vogt90},
infrared \cite{Grenier89,Jandl91}, Rayleigh-Brillouin \cite{Lyons76,Hehlen95,Hehlen95b},
dielectric\cite{Rytz80}, thermal \cite{Salce94} or mechanical \cite{Uwe75}
studies highlighted numerous anomalies in pure KTaO\( _{3} \) crystals at low
temperature. Particularly, Hehlen \textit{et al.} \cite{Hehlen95,Hehlen95b}
observed an extra Brillouin doublet (BD) with a sound-like dispersion relation
in KTaO\( _{3} \) and SrTiO\( _{3} \). In KTaO\( _{3} \), it is seen for
\( T\lesssim  \) 25 K along <001> and <110> directions. This new excitation
seems to appear on top of the quasi-elastic broad central peak (CP) first reported
by Lyons and Fleury \cite{Lyons76}. 

The aim of this paper is to analyse the BD and CP anomalies in pure KTaO\( _{3} \)
in terms of classical phonon kinetics, \textit{i.e.} in terms of phonon-density-fluctuation-induced
light-scattering. In Section \ref{Brillouin}, we present some new Brillouin
measurements studying the anisotropy of the CP and doublets along the three
principal symmetry axes. High-resolution data for low energy dispersion branches
are obtained using inelastic neutron-scattering, as described in Section \ref{Neutrons}.
Experimental results are then used to describe low energy phonon dispersion
surfaces around the BZ centre by mean of a simple phenomenological parameterisation.
Using such a model, we shall show (Section \ref{d2p-2s}) that the anomalous
Brillouin contributions may originate from either two-phonon difference scattering
(TPDS) or second sound (2S) processes, but the observed anisotropy seems to
favour the former.

\section{\textsc{\underbar{\large BRILLOUIN SCATTERING EXPERIMENTS\label{Brillouin}}}\large }

The KTaO\( _{3} \) Ultra High Purity crystals were grown by one of us (LAB)
at ORNL by 'modified spontaneous nucleation' within a slowly cooled flux. The
light-scattering spectra were excited using a single-mode argon-ion laser operating
at 5145 \AA\( \:  \) and observed in backscattering configuration along the
three crystal principal directions. The momentum exchange was \( |\vec{q}| \)
= 5.6 10\( ^{-3} \) \AA\( ^{-1} \). The temperature \( T \) was varied between
5 and 300 K. The spectra were analysed by mean of a six-pass tandem interferometer
that provides an excellent contrast and suppresses the overlap of orders \cite{Sandercock76}.

\begin{figure}
{\par\centering \resizebox*{12.25cm}{8cm}{\includegraphics{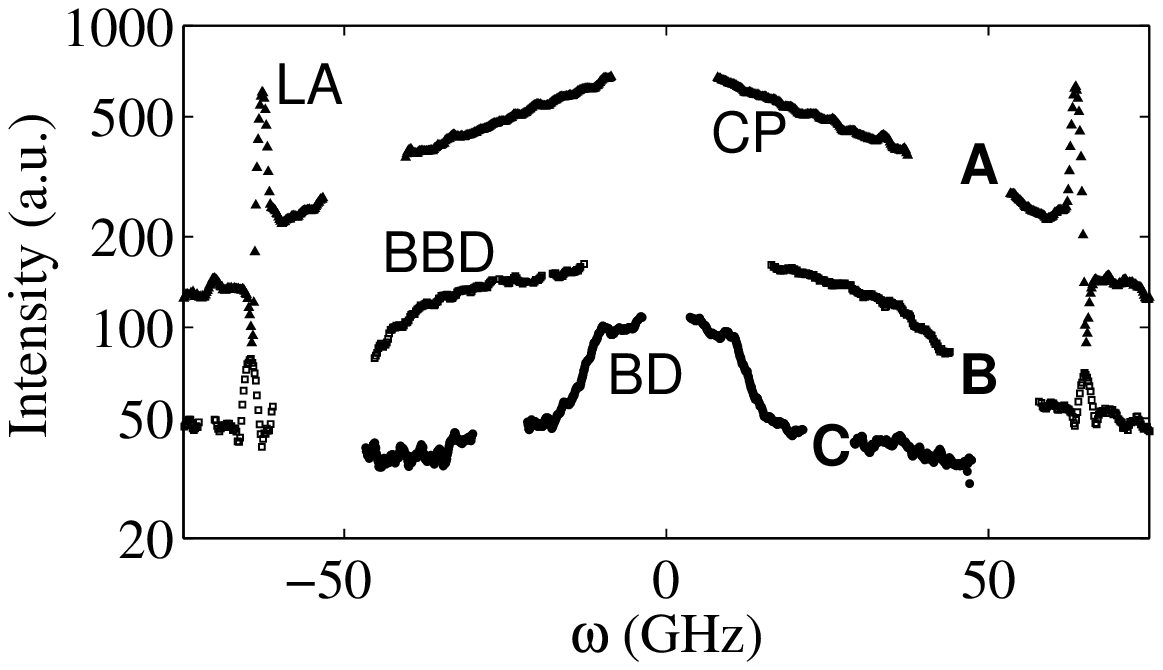}} \par}

\caption{\label{DBL110}Brillouin backscattering spectra of KTaO\protect\( _{3}\protect \)
showing the acoustic LA phonon, central peak (CP), Brillouin Doublet (BD) and
Broad Brillouin Doublet (BBD), measured at \textbf{(A)} 195 K, \textbf{(B)}
31.5 K, and \textbf{(C)} 13 K with \protect\( |\vec{q}|\protect \) // <110>
principal axis. Empty zones correspond to central laser peak and first order
instrumental ghosts \cite{Sandercock76}. Spectra are shifted for clarity. Along
<001> and <111>, one can observe shapes \textbf{A}-\textbf{C} and \textbf{A}-\textbf{B}
respectively.}
\end{figure}

Figure \ref{DBL110} shows typical spectra for various temperatures and \( \vec{q} \)
// <110>. For \( T\gtrsim  \) 100 K, an intense CP appears, strongly coupled
with the LA phonon (curve \textbf{A}). As the temperature is decreased, the
CP contribution becomes weaker, and an unexpected new broad Brillouin doublet
(BBD) grows upon its shoulders (curve \textbf{B}). For \( T\lesssim  \) 25
K, the sound-like BD contribution, first observed by Hehlen \textit{et al.}
\cite{Hehlen95} below 22 K, can be measured (curve \textbf{C}), with a corresponding
speed \( v_{BD}\sim  \) 1100 m/s. It exhibits a width of typically \( \Gamma _{BD}\sim  \)
6 GHz. Along the <001> direction, the BBD excitation is not seen, while along
<111>, no BD appears at temperatures as low as 5 K, and the CP is broader than
along other directions below 100 K \cite{Farhi98}. The CP is clearly coupled
to acoustic phonons along the <110> and <111> directions.
\bigskip{}

\vspace{2cm}

\smallskip{}
\section{\textsc{\underbar{\large LOW ENERGY PHONON DISPERSION CURVES }}\\
\textsc{\underbar{\large AND THEIR PARAMETERISATION\label{Neutrons}}}\large }

In order to analyse the origin and the anisotropy of excitations CP, BD and
BBD in terms of phonon based processes, we parameterised the low energy phonon
dispersion surfaces, which are still significantly populated under 100 K in
KTaO\( _{3} \). A more ambitious modelisation has already been performed along
high symmetry axis \cite{Perry89} using the polarisability model \cite{Perry89,Migoni76}
but it requires both substantial computations and the determination of 15 parameters
in order to be applied just along the three principal directions.

\begin{figure}
{\par\centering \resizebox*{12.25cm}{8cm}{\includegraphics{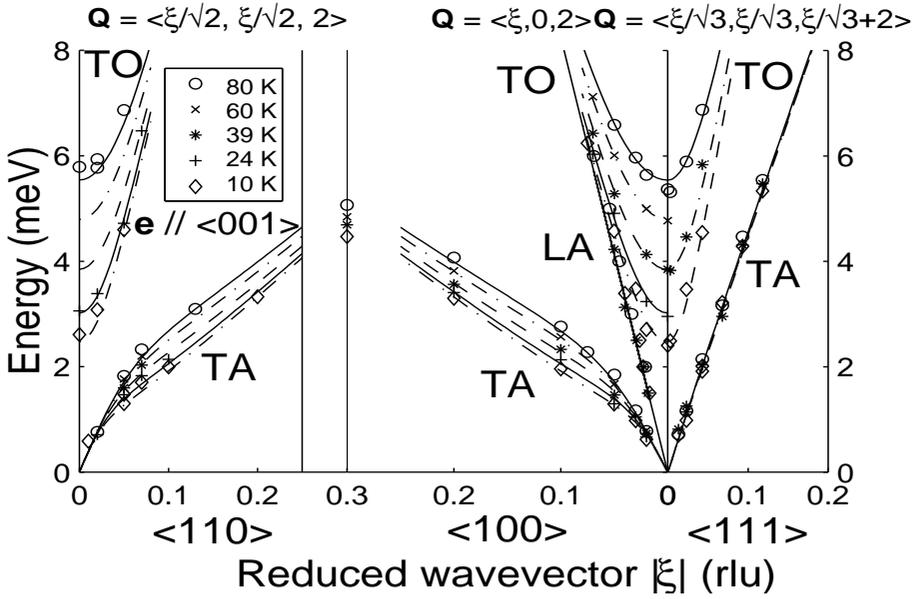}} \par}

\caption{\label{DC}Dispersion curves of KTaO\protect\( _{3}\protect \) low energy
phonons between 10 and 80 K along high symmetry axes. The mean energy uncertainty
is 0.1 meV. The LA mode is for <0 0 2+\protect\( \xi \protect \)>. Lines correspond
to calculations using the Vaks parameterisation.}
\end{figure}

As the precision of the previous studies were not sufficient for our purpose,
we performed new high-resolution inelastic neutron-scattering measurements on
IN14 at the Institut Laue Langevin (ILL), Grenoble, France \cite{Farhi98,Farhi99}.
By taking into account a 4-dimensional modelisation of the instrumental resolution
ellipsoid and of the anisotropic phonon dispersion surfaces, it was possible
to measure phonon branches (Figure \ref{DC}) with a typical energy resolution
of 0.1 meV. This enabled a direct comparison of the soft mode energy (2.6 meV
at 10 K) and damping (0.07 meV at 10 K) with Hyper Raman results near BZ centre
\cite{Vogt90}. Some off-principal axis experiments were also performed. Acoustic
phonon dampings could not be measured, setting a maximum value of about 0.05
meV from instrument resolution.

These Brillouin and neutron-scattering measurements, enabled us to apply the
Vaks parameterisation \cite{Vaks68,Vaks73,Balagurov70} on the five lowest phonon
modes for temperatures below 100 K, in the whole central part of BZ (\( |\vec{q}| \)
< 0.2 rlu) where phonon dispersion surfaces are significantly populated \cite{Farhi98,Farhi99}.
Such a model requires the determination of only 8 parameters, 4 of which can
be measured directly by mean of light-scattering. The resulting curves along
high symmetry axes are shown in Figure \ref{DC}. As presented in Figure \ref{Spectr+D2p}-a,
transverse phonon dispersion surfaces are quite anisotropic, and show deep \emph{'valleys'}
(low energy and group velocities), induced by TA-TO coupling \cite{Axe70},
and \emph{'uphill'} (high energy and group velocities) regions depending on
mode propagation and polarisation directions.

\section{\textsc{\underbar{\large PHONON-DENSITY-FLUCTUATION PROCESSES\label{d2p-2s}}} }

Hehlen \textit{et al.} \cite{Hehlen95b} considered several possible explanations
for the origin of the new BD. The most likely one at the time seemed to be the
onset of second sound (2S), a phenomenon independently predicted in QPE by Gurevich
and Tagantsev \cite{Gurevich88}. The physical reason for this effect is that
the presence of the low frequency TO branches enhances \emph{normal} (momentum
conservation) relaxation processes. This could open the frequency 'window' condition
\( \Gamma _{R}<\omega _{2S}<\Gamma _{N} \) \cite{Hehlen95b,Gurevich88} for
the appearance of 2S near BZ centre, where \( \Gamma _{N} \) and \( \Gamma _{R} \)
are the normal and resistive dampings respectively. In KTaO\( _{3} \), this
excitation is expected to be nearly isotropic in velocity and to couple to acoustic
modes.

Following a thermodynamical approach, it is possible to evaluate the 2S velocity
near BZ centre using \( v_{2S}^{2}=\frac{TS^{2}}{C\cal D} \), where \( S \)
is the entropy density, \( C \) is the heat capacity, and \( \cal D \) is
a density tensor, as detailed in \cite{Gurevich88}. Phonons contributing to
2S should have normal dampings higher than the doublet frequency \( \omega _{BD} \),
in order that the 'window' condition is satisfied. It is possible to evaluate
three-phonon normal dampings \cite{Cowley63} through electrostrictive processes
\cite{Uwe75,Vaks73} using the phonon dispersion surfaces parameterisation.
Numerical computations of this sort indicate that TA dampings are minimal in
<001> and <110> \emph{'valley'} directions, and higher elsewhere, particularly
around \emph{'uphill'} regions \cite{Farhi98}.

Computations of 2S velocity using the Vaks model and the thermalized part of
the phonon spectrum give \( v_{2S} \) = 1100 \( \pm  \) 200 m/s, in excellent
agreement with Brillouin scattering measurements of BD speed. However, it seems
difficult to explain the directional dependence of BD (see Section \ref{Brillouin}
and Figure \ref{DBL110}) in terms of 2S.

\begin{figure}
{\par\centering \resizebox*{6cm}{6cm}{\includegraphics{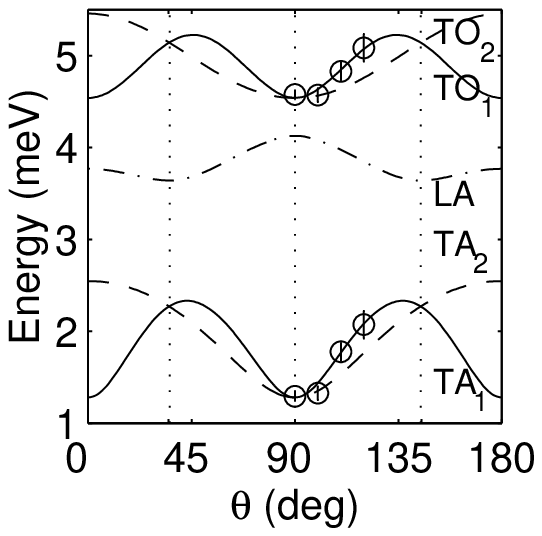}}  \resizebox*{6cm}{6cm}{\includegraphics{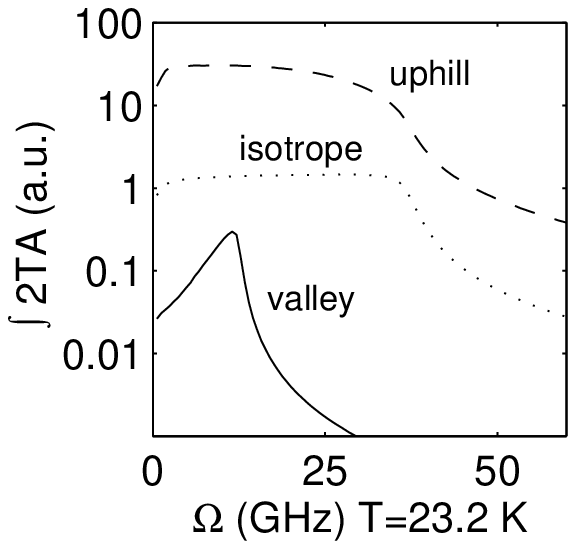}} \par}

\caption{\label{Spectr+D2p}(\textbf{a}) Low energy phonon spectrum at \protect\( T\protect \)
= 10 K from Vaks parameterisation for a rotation \protect\( |\vec{\xi }|\protect \)
= 0.05 rlu. Directions <110>, <111>, and <001> are for angles 0\protect\( ^{o}\protect \)
or 180\protect\( ^{o}\protect \), 35.5\protect\( ^{o}\protect \) or 144.8\protect\( ^{o}\protect \),
and 90\protect\( ^{o}\protect \) respectively. Off principal-axis neutron-scattering
data are also indicated. (\textbf{b}) Computed two-phonon difference scattering
spectrum from different TA model dispersion sheet shapes at \protect\( T\protect \)
= 23 K. Phonon damping is 1 GHz .}
\end{figure}

Second sound is the collective mode of phonon density fluctuation processes
\cite{Wehner72}. When the window condition is not fulfilled (\( \omega _{2S}>\Gamma _{N} \)),
non-thermalized phonon can still interact \emph{via} two-phonon difference scattering
(TPDS) processes. Following Wehner and Klein \cite{Wehner72}, the resulting
Brillouin second order scattering intensity is given by \cite{Farhi98} :

\begin{equation}
\label{eq:ds-line}
S_{TPDS}(\Omega ,q)\propto \sum _{j}\int _{BZ}d^{3}k\frac{k^{n}}{\omega _{\vec{k}}^{2}}N_{\vec{k}}(N_{\vec{k}}+1)\frac{2\Gamma _{\vec{k},j}}{(\Omega -\vec{q}.\vec{v}_{\vec{k},j})^{2}+4\Gamma _{\vec{k},j}^{2}},
\end{equation}
 where \( n \) is 0 and 4 for TO and TA contributions respectively. This expression
has been evaluated for model phonon dispersion surfaces in the case of valley,
isotrope and uphill shapes (see Figure \ref{Spectr+D2p}-b). It is then possible,
in the case of low phonon dampings, to obtain a sharp peak from valley contribution,
or a broader quasi-central spectrum from uphill contributions. The peak or threshold
frequency \( \omega _{TPDS} \) is fixed by the mean group velocity of contributing
non-thermalized phonons, and is smaller in valley than uphill. The TPDS spectrum
broadens with increasing dampings, and then appears as a broad doublet or central
peak, depending on group velocity distribution. Thus, the TPDS spectrum is essentially
anisotropic.

In this scheme, the BD low temperature contribution originates from low damping
and group velocity 'valley' TA phonons TPDS \emph{i.e.} for <001> and <110>
propagations (see Figure \ref{Spectr+D2p}-a). The BBD comes from higher damping
'uphill' TA phonons TPDS \emph{i.e.} for <110> and <111> propagations, at slightly
higher temperatures where those regions are significantly populated. The BD
and BBD contributions disappear when the soft mode energy increases, and the
TA-TO coupling induced valleys open up. For the TO modes, with high dampings,
a very broad quasi-central peak is expected, corresponding to the CP contribution.

Wehner and Klein \cite{Wehner72} predicted that in various crystals, including
SrTiO\( _{3} \), the integrated intensities of phonon density fluctuation processes
(i.e. 2S or TPDS) could be 10 to 100 times the value given by the Landau-Placzek
ratio. It is then quite plausible that these intensities become comparable,
or even higher than phonon light scattering intensities in KTaO\( _{3} \),
which would thus bring them in agreement with observed intensities.

\section{\underbar{\large CONCLUSION}\large }

We have shown that the phonon spectrum and dampings in KTaO\( _{3} \) are consistent
with two phonon-density-related scenarios for the appearance of quasi-elastic
central peak and doublets. One scenario is related to temperature waves (second
sound) involving a part of the spectrum that can be thermalized during the doublet
lifetime \cite{Hehlen95b,Gurevich88}. The other scenario is related to the
two-phonon difference scattering of non-thermalized transverse phonons from
phonon sheet 'valleys' or 'uphill' regions \cite{Farhi98}. The position of
the sharp doublet BD is consistent with both scenarios whereas the anisotropy
of the scattering favours the second one.


\begin{thebibliography}{10}
\bibitem{Kurtz75}{\small K. Kurtz,} \underbar{\small Trans. Am. Cryst. Assoc.} \textbf{\small 2}{\small ,
63 (1975).}{\small \par}
\bibitem{Muller79}{\small K.A. M\"{u}ller and H. Burkard,} \underbar{\small Phys. Rev. B} \textbf{\small 19}{\small ,
3593 (1979).}{\small \par}
\bibitem{Tosatti94}{\small E.  Tosatti  and  R. Marto\~{n}\`{a}k,} \underbar{\small Sol. Stat.
 Comm.} {\small } \textbf{\small 92}{\small , 167 (1994). }{\small \par}
\bibitem{Unoki67}{\small H. Unoki and T. Sakudo,} \underbar{\small J. Phys. Soc. Jap.} \textbf{\small 23}{\small ,
546 (1967). }{\small \par}
\bibitem{Thomas68}{\small H. Thomas and K.A. M\"{u}ller ,} \underbar{\small Phys. Rev.  Lett.}
{\small } \textbf{\small 21}{\small ,  1256 (1968).}{\small \par}
\bibitem{Shirane69}{\small G. Shirane and Y. Yamada,} \underbar{\small Phys. Rev.} \textbf{\small 177}{\small ,
858 (1969).}{\small \par}
\bibitem{Axe70}{\small J.D. Axe, J. Harada, and G. Shirane,} \underbar{\small Phys. Rev. B}
\textbf{\small 1}{\small , 1227 (1970). }{\small \par}
\bibitem{Comes72}{\small R. Com\`{e}s and G. Shirane,} \underbar{\small Phys. Rev. B} \textbf{\small 5}{\small ,
1886 (1972). }{\small \par}
\bibitem{Perry89}{\small C.H.  Perry,  R.  Currat,  H.  Buhay,  R.M.  Migoni,  W.G. Stirling,
and J.D. Axe,} \underbar{\small Phys. Rev. B} \textbf{\small 39}{\small , 8666
(1989). }{\small \par}
\bibitem{Vogt90}{\small H. Vogt and H. Uwe,} \underbar{\small Phys. Rev. B} \textbf{\small 29}{\small ,
1030 (1984).} \emph{\small See also} {\small H. Vogt,} \underbar{\small Phys.
Rev. B} \textbf{\small 41}{\small , 1184 (1990).}{\small \par}
\bibitem{Grenier89}{\small P. Grenier, G. Bernier, S. Jandl, B. Salce, and L.A. Boatner,} \underbar{\small J.
Phys. : Cond. Matt.} \textbf{\small 1}{\small , 2515 (1989). }{\small \par}
\bibitem{Jandl91}{\small S.  Jandl, M. Banville, P. Dufour, S. Coulombe,  and  L.A. Boatner,}
\underbar{\small Phys. Rev. B} \textbf{\small 43}{\small , 7555, (1991).}{\small \par}
\bibitem{Lyons76}{\small K.B.  Lyons and P.A. Fleury,} \underbar{\small Phys. Rev.  Lett.} {\small } \textbf{\small 37}{\small ,
 161 (1976).}{\small \par}
\bibitem{Hehlen95}{\small B.  Hehlen,} \underbar{\small Doctoral Thesis}{\small , Universit\'{e}
Montpellier II (1995). }{\small \par}
\bibitem{Hehlen95b}{\small B. Hehlen, A-L. P\'{e}rou, E. Courtens, and R. Vacher,} \underbar{\small Phys.
Rev. Lett.} \textbf{\small 75}{\small , 2416 (1995). }{\small \par}
\bibitem{Rytz80}{\small D. Rytz, U.T. H\"{o}chli, and H. Bilz,} \underbar{\small Phys. Rev.
B} \textbf{\small 22}{\small , 359 (1980). }{\small \par}
\bibitem{Salce94}{\small B. Salce, J.L. Gravil, and L.A. Boatner,} \underbar{\small J. Phys.
:  Cond. Matt.} \textbf{\small 6}{\small , 4077 (1994). }{\small \par}
\bibitem{Uwe75}{\small H. Uwe and T. Sakudo,} \underbar{\small J. Phys. Soc. Jap.} \textbf{\small 38}{\small ,
183 (1975). }{\small \par}
\bibitem{Sandercock76}{\small J. Sandercock,} \underbar{\small J. Phys. E} \textbf{\small 9}{\small ,
566 (1976).}{\small \par}
\bibitem{Farhi98}{\par\raggedright {\small E. Farhi,} \underbar{\small Doctoral Thesis}{\small ,
Universit\'{e} Montpellier II (1998). <http://www.ldv.univ-montp2.fr:7082/\~{}manuf/these.html>}\small \par}
\bibitem{Migoni76}{\small R. Migoni, H. Bilz, and D. B\"{a}uerle,} \underbar{\small Phys. Rev.
Lett.} \textbf{\small 25}{\small , 1155 (1976). }{\small \par}
\bibitem{Farhi99}{\small E. Farhi, A.K. Tagantsev, R. Currat, B. Hehlen, L.A. Boatner, and} {\small E.
Courtens, submitted to} \underbar{\small Eur. Phys. Jour. B} {\small (1999).}{\small \par}
\bibitem{Vaks68}{\small V.G. Vaks,} \underbar{\small Zh. Eksp. Teor. Fiz.} \textbf{\small 54}{\small ,
910 (1968) {[}}\underbar{\small Sov. Phys. JETP} \textbf{\small 27}{\small ,486
(1968){]}. }{\small \par}
\bibitem{Vaks73}{\small V.G. Vaks,} \underbar{\small Vvedenie v Mikroskopicheskuyu Teoriyu Segnetoelectrikov}
{\small {[}Introduction to the Microscopic Theory of Ferroelectrics{]}, Nauka,
Moscow (1973). }{\small \par}
\bibitem{Balagurov70}{\small B.Y. Balagurov, V.G. Vaks and B.I. Shklovshii,} \underbar{\small Zh.
Eksp. Teor. Fiz.} \textbf{\small 12}{\small , 89 (1970) {[}}\underbar{\small Sov.
Phys.-Solid State} \textbf{\small 12}{\small , 70 (1970){]}. }{\small \par}
\bibitem{Gurevich88}{\small V.L. Gurevich and A.K. Tagantsev,} \underbar{\small Zh. Eksp. Teor.
Fiz.} \textbf{\small 94}{\small , 370 (1988) {[}}\underbar{\small Sov. Phys.
JETP} \textbf{\small 67}{\small , 206 (1988){]}.}{\small \par}
\bibitem{Cowley63}{\small R.A. Cowley,} \underbar{\small Adv. Phys.} \textbf{\small 12}{\small ,
421 (1963).}{\small \par}
\bibitem{Wehner72}{\small R.K.Wehner and R. Klein,} \underbar{\small Physica (Utrecht)} \textbf{\small 62}{\small ,
161 (1972).}{\small \par}
\end{thebibliography}
\end{document}